\documentclass[showpacs,twocolumn,aps,pra,longbibliography,superscriptaddress,notitlepage]{revtex4-2}
\usepackage{siunitx}
\usepackage{graphicx}
\usepackage{physics}
\usepackage{wrapfig}
\usepackage{changepage}
\usepackage{amsmath,amssymb,amsthm,mathrsfs,amsfonts,dsfont,mathtools}
\usepackage[caption=false]{subfig}
\usepackage{epsfig}
\usepackage{color,xcolor}
\usepackage{adjustbox}
\usepackage{tabularx}
\usepackage{array}
\usepackage{multirow}
\usepackage{tabu}
\usepackage{bm}
\usepackage{enumerate}
\usepackage[colorlinks=true,linkcolor=blue,citecolor=blue,urlcolor=blue,pdfauthor={ },pdftitle={ },pdfsubject={ },pdfkeywords={ }]{hyperref}
\usepackage{newtxtext,newtxmath}
\usepackage{makecell} 
\usepackage{url}
\usepackage{booktabs}
\usepackage{nicefrac} 
\usepackage{makeidx}  
\usepackage{leftindex}
\usepackage[ruled,vlined,linesnumbered]{algorithm2e}
\usepackage{cleveref}
\usepackage{mwe}

\usepackage{comment}
\usepackage{xurl}

\usepackage{color}
\definecolor{dred}{rgb}{.8,0.2,.2}
\definecolor{ddred}{rgb}{.8,0.5,.5}
\definecolor{dblue}{rgb}{.2,0.2,.8}
\definecolor{dgreen}{rgb}{.2,0.5,.2}

\begin{document}
\title{Diabatic quantum annealing for training energy-based generative models}

\author{Gilhan Kim}
\thanks{These authors contributed equally to this work.}
\affiliation{Department of Statistics and Data Science, Yonsei University, Seoul 03722, Republic of Korea}

\author{Ju-Yeon Gyhm}
\thanks{These authors contributed equally to this work.}
\affiliation{Department of Physics and Astronomy, Seoul National University, Seoul 08826, Republic of Korea}

\author{Daniel K. Park}
\email{dkd.park@yonsei.ac.kr}
\affiliation{Department of Statistics and Data Science, Yonsei University, Seoul 03722, Republic of Korea}
\affiliation{Department of Applied Statistics, Yonsei University, Seoul 03722, Republic of Korea}
\affiliation{Department of Quantum Information, Yonsei University, Seoul 03722, Republic of Korea}

\begin{abstract}
Energy-based generative models, such as restricted Boltzmann machines (RBMs), require unbiased Boltzmann samples for effective training. Classical Markov chain Monte Carlo methods, however, converge slowly and yield correlated samples, making large-scale training difficult. We address this bottleneck by applying the analytic relation between annealing schedules and effective inverse temperature in diabatic quantum annealing. By implementing this prescription on a quantum annealer, we obtain temperature-controlled Boltzmann samples that enable RBM training with faster convergence and lower validation error than classical sampling. We further identify a systematic temperature misalignment intrinsic to analog quantum computers and propose an analytical rescaling method that mitigates this hardware noise, thereby enhancing the practicality of quantum annealers as Boltzmann samplers. In our method, the model’s connectivity is set directly by the qubit connectivity, transforming the computational complexity inherent in classical sampling into a requirement on quantum hardware. This shift allows the approach to extend naturally from RBMs to fully connected Boltzmann machines, opening opportunities inaccessible to classical training methods.
\end{abstract}

\maketitle


\section{Introduction} Restricted Boltzmann machines (RBMs) stand as a paradigmatic model at the interface of statistical physics and machine learning, demonstrating how physical principles underpin modern machine learning and artificial intelligence~\cite{Hinton2002,Smolensky1986}. Training RBMs requires statistically correct samples from the Boltzmann distribution defined by the model’s energy function. Inaccurate sampling leads to biased gradient estimates, slower convergence, and degraded generalization. As a result, sampling has long been recognized as the principal bottleneck in scaling RBMs and related energy-based models. 

Classical approaches rely on Markov chain Monte Carlo (MCMC) methods~\cite{Newman1999} such as contrastive divergence (CD) and persistent contrastive divergence (PCD)~\cite{Hinton2002,Tieleman2008}, which converge slowly and generate strongly correlated samples. Thus, producing large numbers of independent samples within practical time scales remains a major challenge. Overcoming this inefficiency is critical not only for making RBMs practical at scale, but also for advancing the broader role of physics-inspired models in modern machine learning.


Analog quantum computers~\cite{Kadowaki1998,farhi2000quantum} have been proposed as hardware-based samplers capable of generating Boltzmann-like distributions through quantum dynamics~\cite{Adachi2015,AminPRA2015}. Early demonstrations showed qualitative success but suffered from a critical drawback: the effective sampling temperature was not properly controlled, preventing reliable alignment between hardware samples and the model distribution required for training. Subsequent studies often treated the sampling temperature as an empirical fitting parameter, adjusting it from histograms at each training epoch~\cite{BenedettiPRA2016,korenkevych2016benchmarking}. This ad hoc procedure compromises reproducibility and blurs the distinction between genuine quantum dynamics and artifacts of arbitrary parameter fitting.

Recent theoretical work on diabatic quantum annealing (DQA) has revealed an analytic relation between the annealing schedule and an effective inverse temperature $\beta_{\mathrm{integral}}$~\cite{DQApaper}. This result implies that, given a schedule of time dependent Hamiltonian, the temperature of the resulting samples can be prescribed in advance rather than inferred a posteriori. The relation provides a principled way to align hardware sampling with the conventions of Boltzmann-machine training, ensuring that the model distribution is preserved rather than distorted.

Here we apply this theoretical prescription, for the first time, to practical machine learning: Boltzmann sampling for faster and more accurate generative modeling.
By implementing DQA with controlled annealing schedules, we generate calibrated Boltzmann samples and use them to train RBMs. The quantum-assisted parameter updates demonstrably outperform classical MCMC sampling, achieving faster convergence and lower validation error under identical model conditions. Beyond this demonstration, we identify a systematic source of temperature misalignment in the quantum annealer: the effective sampling temperature deviates from the theoretical value due to thermal effects, noise, and control imperfections. Crucially, we provide a method to correct these deviations through analytic rescaling, thereby restoring the intended sampling behavior and enabling principled training. Together, these results establish quantum annealers as practical Boltzmann samplers and highlight error-mitigation strategies essential for physics-based machine learning on quantum hardware.

\begin{figure}
\includegraphics[width=\columnwidth]{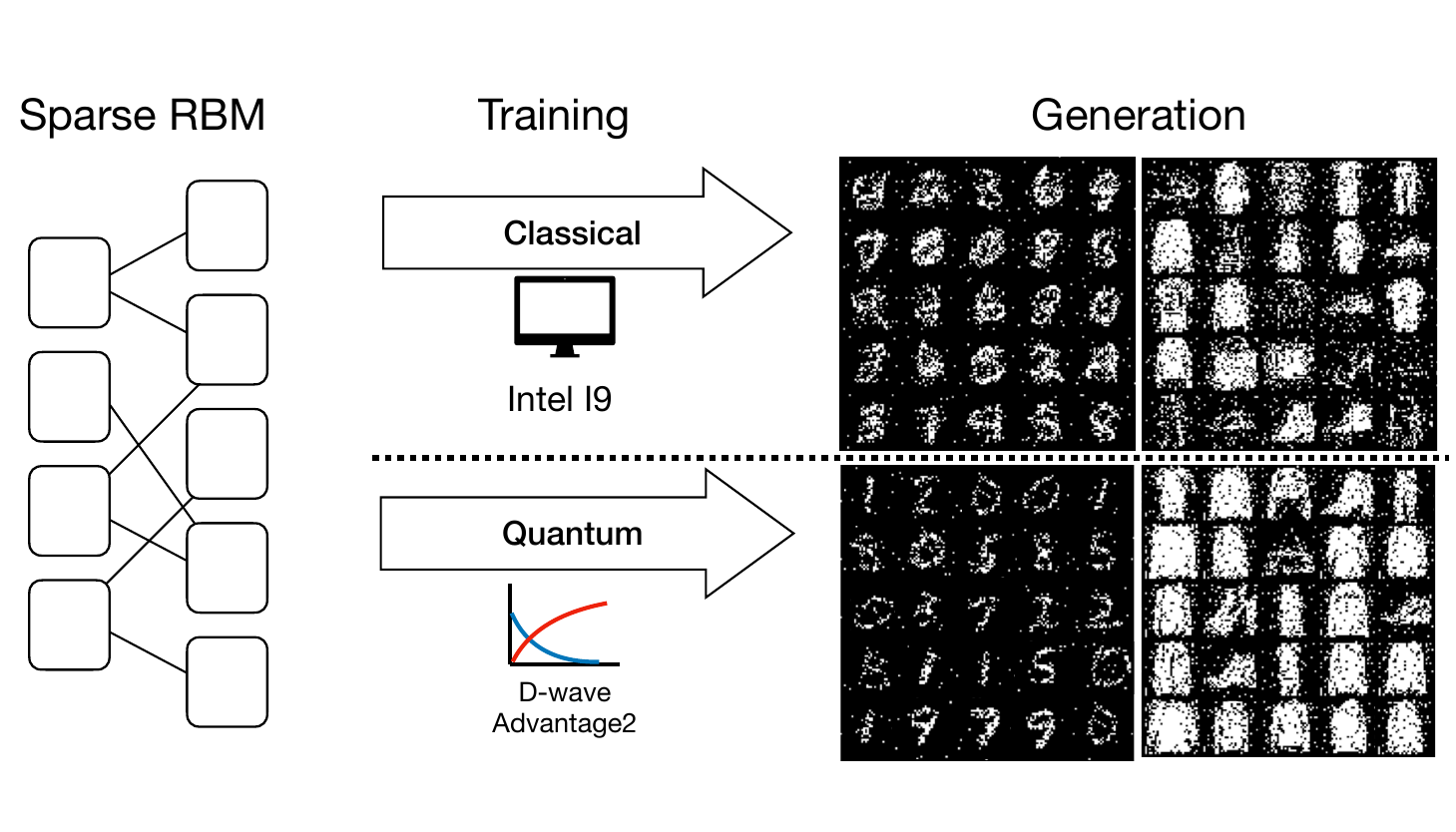}
\caption{\label{fig:main_fig} Schematic comparison of RBM training using classical persistent contrastive divergence (PCD) and diabatic quantum annealing (DQA)–based sampling. In both cases, the same RBM architecture and training procedure are used; the only difference lies in the sampling method employed to estimate model expectations during learning. The RBM consists of 784 visible units and 1200 hidden units, with each node connected on average to 18.17 others (standard deviation 2.37). The right panels show representative samples generated from the trained RBM (left: MNIST, right: Fashion-MNIST), illustrating the quality of the learned generative model under each sampling method.
}
\end{figure}

\section{Training RBMs via DQA} An RBM is a stochastic neural network consisting of one visible layer with $N_v$ nodes and one hidden layer with $N_h$ nodes. Let $v\in\{\pm1\}^{N_v}$ and $h\in\{\pm1\}^{N_h}$. The model parameters are encoded in the weight matrix $J\in\mathbb{R}^{N_v\times N_h}$, where each element $J_{ij}$ specifies the coupling strength between visible node $i$ and hidden node $j$.
The RBM energy is
\begin{equation}\label{eq:energy}
E(v,h) = - v^\top J h,
\end{equation}
which defines the joint probability distribution $Q(v,h) = \exp[-\beta E(v,h)]/Z(\beta)$, where $Z(\beta)$ denotes the partition function at inverse temperature $\beta$. 
The training process minimizes the Kullback-Leibler (KL) divergence between target distribution $P$ and model's marginal distribution $Q_V=\sum_h Q(v,h)$,
\begin{equation}\label{eq:update}
D_\mathrm{KL}(P\|Q_V)=\sum_v P(v)\log \frac{P(v)}{Q_V(v)}.
\end{equation} 
The gradient of the KL divergence can be calculated using the training dataset $\mathcal{D}$ and the model-sampled set $\mathcal{M}$:
\begin{equation}\label{eq:update}
\partial_{J_{ij}} \mathcal{L} \;=\; \big\langle v_i h_j^\top \big\rangle_{\mathcal{D}} \;-\; \big\langle v_i h_j^\top \big\rangle_{\mathcal{M}},
\end{equation}
where $\left<...\right>_{\mathcal{A}}$ denotes the expectation value over dataset $\mathcal{A}$.   
In classical training, the calculation of $\left<...\right>_\mathcal{M}$ relies on samples obtained from Markov chains, typically implemented with contrastive divergence (CD) or persistent contrastive divergence (PCD) algorithm. However, MCMC approach is slow to converge and produces correlated samples, making large-scale RBM training practically infeasible.

To overcome this bottleneck, we replace the classical sampling procedure with diabatic quantum annealing (DQA). In DQA, the system evolves under a time-dependent Hamiltonian,
\begin{equation}
H(t)=A(t)\,H_\mathrm{mixing} + B(t)\,H_\mathrm{problem},\quad t\in[0,\tau],
\end{equation}
where $H_\mathrm{problem}$ encodes the RBM energy function that defines the target distribution, and $A(t)$ and $B(t)$ are the annealing schedule functions. For relatively short annealing times $\tau$, non-adiabatic transitions generate a distribution that approximates a Boltzmann distribution with effective inverse temperature $\beta$~\cite{DQApaper}. A key advantage of this approach is that each annealing run yields an independent sample, avoiding the correlations inherent in Markov chains.

The central idea is to identify annealing schedules and durations that yield an effective inverse temperature of $\beta=1$, thereby preserving the shape of the target Boltzmann distribution. For a problem Hamiltonian $H_\mathrm{problem}$ containing only two-body interaction, the effective inverse temperature is given analytically by
\begin{align}\label{eq:DQAintegral}
    \beta_{\mathrm{integral}}
    \;=\;
    2 \int_{0}^{\tau}\! dt\, B(t)\,
    \sin\!\left[\,4\int_{t}^{\tau}\! ds\, A(s)\right],
\end{align}
which directly links the control schedule to the effective temperature and enables the determination of these conditions without empirical fitting~\cite{DQApaper}.
This relation holds in the small-$\beta E$ regime, which can be satisfied within the native energy scale of the hardware and the effective inverse temperature realized during fast annealing. The resulting samples are used to evaluate the model expectation $\left<...\right>_\mathcal{M}$ in Eq.~(\ref{eq:update}), which drives gradient-based parameter updates during RBM training. The learning dynamics are thus driven by Boltzmann samples that faithfully reflect the intended distribution.

Specifically, in this study we confirm that using the fast annealing mode of the D-Wave Advantage2 machine with the default schedule~\cite{dwave-schedule} and an annealing time of 5 ns yields the effective inverse temperature of $\beta = 1.5$. We adopt this value in the sampling process. In this setting, each annealing run produces one spin configuration sampled from the Boltzmann distribution with to $\beta\simeq 1$.

To benchmark the approach, we trained RBMs under two settings: one using PCD as the sampler, and the other using DQA samples obtained from the D-Wave device. The training procedure is detailed in Algorithm~\ref{alg:training}. The total number of training epochs is denoted as $T$, and the number of samples per epoch is $S$. We set $T=20$ and $S=3000$. For DQA, an additional parameter $\alpha$ (see Eq.~(\ref{eq:alpha})) is required for compensating for unwanted device effects, which will be explained in the next section. For PCD, an additional parameter $K$ specifies the number of decorrelation steps used to reduce correlations between successive samples. From the perspective of statistical physics, $K$ scales as $K\sim(\text{system size})^z$, where $z$ is the dynamical critical exponent (e.g., about $2.0$ for Ising-type models~\cite{hohenberg1977theory}), implying polynomial growth with system size. In practice, however, such scaling is rarely implemented; instead, small constant values such as $K=5$ or $10$ are commonly used. In our experiments, we likewise fixed $K$ to a constant value for practicality, but chose a larger value ($K=100$) to better suppress correlations.

\begin{algorithm}[t]
\caption{Training of RBMs via Quantum (DQA) or Classical (PCD) Sampling}
\label{alg:training}
 Input \ {$T$, $S$, $K$, $\alpha$}
\DontPrintSemicolon

\If{use DQA}{
  \For{$t \gets 1$ \KwTo $T$}{
    Embed $\{J_{ij}/\alpha\}$ onto the D\mbox{-}Wave machine\;
    \For{$s \gets 1$ \KwTo $S$}{
      Run DQA and store the states $(v, h)$ to $\mathcal{M}$\;
    }
    $J_{ij} \gets J_{ij} + \eta\Big(
      \langle v_i h_j^\top \rangle_{\mathrm{data}}
      - \langle v_i h_j^\top \rangle_{\mathcal{M}}
    \Big)$ for all $(i,j)$\;
  }
}
\If{use PCD}{
  \For{$t \gets 1$ \KwTo $T$}{
    Initialize random hidden state $h$\;

    \For{$s \gets 1$ \KwTo $S$}{
      \For{$k \gets 1$ \KwTo $K$}{
        Sample $v$ from $h$\;
        Sample $h$ from $v$\;
      }
      Save the states $(v,h)$ to $\mathcal{M}$\;
    }
    $J_{ij} \gets J_{ij} + \eta\Big(
      \langle v_i h_j^\top \rangle_{\mathrm{data}}
      - \langle v_i h_j^\top \rangle_{\mathcal{M}}
    \Big)$ for all $(i,j)$\;
  }
}
\end{algorithm}

\begin{figure}
\includegraphics[width=\columnwidth]{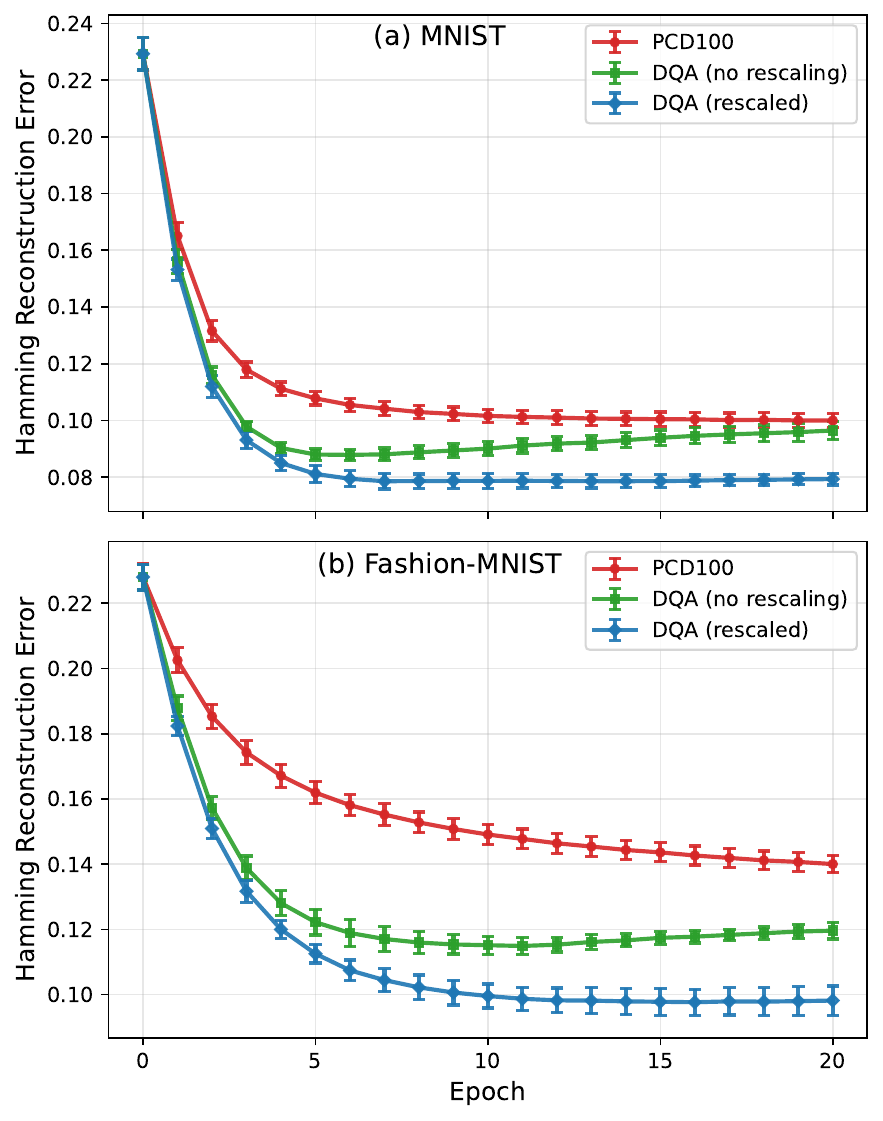}
\caption{\label{fig:val_fig}Validation errors during RBM training on (a) MNIST and (b) Fashion-MNIST. Curves show mean Hamming reconstruction error over 10 independently trained RBMs, with error bars indicating one standard deviation. The red curves correspond to classical PCD training, while the green and blue curves show DQA-based training before and after applying the parameter-rescaling method that corrects for temperature misalignment, respectively. The rescaled DQA samples yield consistently lower reconstruction error, demonstrating the importance of correcting hardware-induced temperature misalignment.
}
\end{figure}

As an illustrative test case, we used MNIST~\cite{lecun2010mnist} and Fashion-MNIST~\cite{xiao2017fashionmnist}. In both datasets, the DQA-trained RBM converged faster and achieved lower reconstruction error than the PCD baseline as illustrated in Fig.~\ref{fig:val_fig}. Moreover, the end-to-end generative performance, measured by sample quality, was consistently better with quantum-assisted training. Fig.~\ref{fig:main_fig} presents the schematic results of this study. These results demonstrate that DQA provides a practical sampling engine for RBM training, improving both training efficiency and model performance. 

We compared the average time required to obtain a single sample per epoch. In PCD training, the inner loop with $K=100$ Gibbs steps requires approximately \num{1.63e-2} seconds per sample. By contrast, in DQA training, the sampling time is about \num{2.56e-4} seconds. Thus, DQA provides a sampling procedure that is roughly 64 times faster than PCD with $K=100$. 

We further examined how the advantage of DQA-based training scales with system size. As discussed in Appendix Section A, DQA consistently exhibits faster convergence and achieves lower validation error as the model dimension increases, indicating improved sampling fidelity, with the performance gap relative to classical PCD widening at larger hidden-layer sizes within the range accessible to current hardware. Appendix Section B analyzes the system-size scaling of the sampling time and shows that classical PCD incurs an additional correlation-induced overhead, whereas DQA replaces this with a static state-preparation cost. This qualitative difference is captured by the asymptotic scaling relations between the total system size $N=N_h+N_v$ and the time required to obtain one sample: 
\begin{equation}
\begin{aligned}
t_{\mathrm{PCD}} &\in \mathcal{O}\!\left(N^{2+z}\right)\text{ with } z>0, \\
t_{\mathrm{DQA}} &\in \mathcal{O}\!\left(N^{2}\right),
\end{aligned}
\end{equation}


\section{Temperature calibration} For practical and reliable applications of DQA to RBM training, it is critical to validate that the samples produced by the quantum hardware follow the target Boltzmann distribution.
Previous studies have established that the D-Wave machine reproduces key physical properties expected from coherent quantum dynamics~\cite{king2023quantum,king2018observation,sathe2025classical}. Accordingly, our focus here is on the reproducibility of the effective sampling temperature. To this end, we compared energy histograms obtained from the device with those predicted by theory and numerical simulations of the programmed two-level Hamiltonian under the same annealing schedule~\cite{dwave-schedule}.

\begin{figure}
\includegraphics[width=\columnwidth]{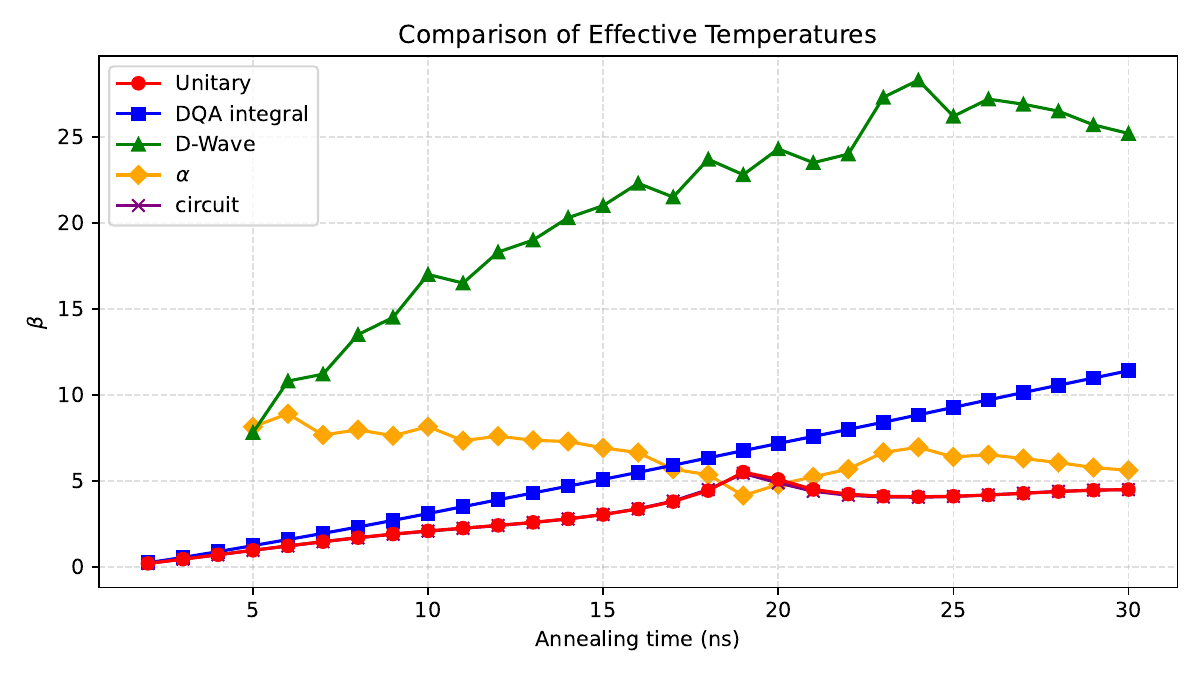}
\caption{\label{fig:2level_fig}Inverse temperature obtained from unitary simulation (red circles), from the integral expression in Eq.~(\ref{eq:DQAintegral}) (blue squares), and empirically estimated from the D-Wave device (green triangles). The rescaling factor $\alpha$ (yellow diamonds) is defined in Eq.~(\ref{eq:alpha}). The purple curve shows a discretized, Trotterized simulation of the annealing dynamics performed using Qiskit. While $\beta_{\mathrm{dwave}}$ deviates from the theoretical predictions, $\alpha$ remains within the range $5$–$7$ over the annealing-time window examined.
}
\end{figure}

Figure~\ref{fig:2level_fig} shows the effective inverse temperatures $\beta$ obtained by different methods in a two-level system evolved under the default D-Wave annealing schedule~\cite{dwave-schedule}. The reference value $\beta_\mathrm{unitary}$ (red circle), obtained by solving the Schr\"{o}edinger equation numerically, serves as the ground truth. The analytic estimate $\beta_\mathrm{integral}$ (blue square), derived directly from Eq.(\ref{eq:DQAintegral}) without empirical fitting, determines the annealing time corresponding to $\beta \simeq 1$, which is then adopted for RBM training. In the small-$\beta$ regime, it closely approximates the ground truth. Finally, $\beta_\mathrm{dwave}$ (green triangles) is obtained from D-Wave samples empirically via
\begin{align}
\beta_{\mathrm{dwave}}
= \frac{1}{E_1-E_0}\,\ln\!\left(\frac{\hat p_0}{\hat p_1}\right),
\end{align}
where $E_0$ and $E_1$ are the ground and the first excited state energies, respectively, and $\hat p_0$ and $\hat p_1$ denote the observed state probabilities.

Our analysis reveals a systematic deviation in the effective temperature as depicted in Fig.~\ref{fig:2level_fig}. The empirical inverse temperature $\beta_{\mathrm{dwave}}$ is consistently larger than both the theoretical value derived from Eq.~(\ref{eq:DQAintegral}) (blue square) and the unitary simulation (red circle). Within the annealing-time range examined, the ratio of $\beta_{\mathrm{dwave}}$ and $\beta_{\mathrm{unitary}}$ fluctuates between approximately 5 and 7. This implies that the hardware generates distributions corresponding to a lower effective temperature than intended. 
The discrepancy is consistent across different annealing times, suggesting that it arises from uncontrollable hardware effects rather than finite-sampling noise. 

To compensate for this misalignment, we rescale the programmed energy function by a factor that aligns the effective temperature of the device with the target value. Specifically, this is implemented by dividing the coupling strengths by the ratio,
\begin{align}\label{eq:alpha}
    \alpha:=\frac{\beta_{\mathrm{dwave}}} { \beta_{\mathrm{unitary}}}
       ,
\end{align}
where all inverse temperatures are evaluated at the fixed annealing time used
for RBM training. This rescaling is applied prior to embedding the problem on the quantum annealer (see Algorithm~\ref{alg:training} line 4). This adjustment mitigates hardware imperfections and noise, bringing the resulting distribution into closer agreement with the target Boltzmann distribution. In this study, we adopt $\alpha=7$ for $\tau=5 \;\mathrm{ns}$, corresponding to the operating point used to generate the results shown in Fig.~\ref{fig:val_fig}.

We validated this calibration by training RBMs with and without rescaling. The rescaled DQA produced significantly higher-quality samples (compare Fig.~\ref{fig:main_fig} green and blue lines). The corrected distributions yield more accurate thermal expectation values, which in turn lead to better gradient estimates and faster convergence of the RBM training. These findings establish temperature rescaling as an essential step for harnessing quantum annealers as practical Boltzmann samplers for machine learning.

\section{Conclusion and discussion} We have demonstrated the first experimental application of the analytic relation between annealing schedules $\tau$ and effective inverse temperature $\beta$ in diabatic quantum annealing to the training of restricted Boltzmann machines. By employing schedule-controlled sampling on a quantum annealer, we generated Boltzmann distributions at prescribed temperatures, trained RBMs, and showed that quantum-assisted parameter updates outperform classical MCMC sampling in both convergence speed and validation error. We further identified a temperature misalignment inherent to analog quantum computing platforms and demonstrated that it can be systematically corrected by analytic rescaling, thereby restoring principled learning dynamics. While our demonstration used the superconducting-based D-Wave machine, the same framework is directly applicable to other quantum annealing platforms, including neutral-atom systems~\cite{kim2022rydberg} such as those under active development by Pasqal~\cite{henriet2020quantum} and QuEra~\cite{wurtz2023aquila}.

Beyond this demonstration, our results carry several broader implications. First, unlike many quantum machine learning studies that train RBMs on a significantly lower-dimensional representation of MNIST or similar datasets (obtained via dimensionality reduction)~\cite{Wilson2021,Dixit2022,Dixit2021}, our work trains directly on the full $28\times28$ pixel images without dimensionality reduction or volume reduction. The number of qubits in this study is 1984 (784 visible + 1200 hidden), setting a new benchmark for the scale of quantum machine learning experiments.

Second, while RBMs adopt a bipartite structure to make Gibbs sampling tractable, DQA-based sampling does not impose such a restriction. This opens the possibility of reviving the more general Boltzmann machines (BMs)~\cite{Hinton1983}, whose unrestricted connectivity can better capture correlations and structural priors but has been largely abandoned due to sampling intractability. In our approach, the model connectivity is dictated directly by the qubit connectivity of the annealer, shifting the challenge from algorithmic complexity to hardware realization.

Third, our current experiments are constrained by the sparse qubit connectivity of present-day hardware, such as the Zephyr topology of D-Wave~\cite{zephyr}. As quantum annealers evolve toward denser and larger-scale connectivity, the same principles can extend to richer models. In particular, quantum-assisted sampling could enable the training of more expressive generative models such as variational autoencoders (VAEs)~\cite{kingma2013auto}, which also require sampling for parameter updates, thereby broadening the role of analog quantum computers as practical tools for generative modeling.


Finally, the quantum process studied here can also be implemented on gate-based quantum circuits by approximating the annealing schedule with piecewise-constant controls and applying Trotterization~\cite{trotter1959product}. We verified this correspondence in Qiskit~\cite{Qiskit} simulations, as illustrated by the purple curve ($\times$ markers) in Fig.~\ref{fig:2level_fig}, which closely reproduces the exact unitary dynamics over the annealing-time range considered. At present, the number of physical qubits available on gate-based hardware is smaller than on quantum annealers, making large-scale tasks such as full MNIST training not yet available. However, circuit-based platforms have a clear path toward scalability through the theory of quantum error correction and fault-tolerance~\cite{aharonov1997fault,shor1995scheme,terhal2015quantum,fowler2012surface}. Moreover, the extensive toolbox of error mitigation techniques developed for gate-based devices~\cite{temme2017error,endo2018practical,bravyi2021mitigating,QMEM_Lee_2023,Qutility} can be directly applied in this framework. Extending the gate-based implementation to larger systems and more general circuit constructions, and developing an analytic understanding of how Trotterization and hardware imperfections affect the effective temperature relation, remain important directions for future work. Another interesting avenue is to explore whether variational tuning of circuit parameters could further improve sampling quality.

\textit{Acknowledgments}---This work is supported by Institute of Information \& communications Technology Planning \& evaluation (IITP) grant funded by the Korea government (No. 2019-0-00003, Research and Development of Core Technologies for Programming, Running, Implementing and Validating of Fault-Tolerant Quantum Computing System), the National Research Foundation of Korea (RS-2023-NR119931,RS-2025-02309510), the Ministry of Trade, Industry, and Energy (MOTIE), Korea, under the Industrial Innovation Infrastructure Development Project (RS-2024-00466693), and by Korean ARPA-H Project through the Korea Health Industry Development Institute (KHIDI), funded by the Ministry of Health \& Welfare, Korea (RS-2025-25456722).

%

\clearpage
\onecolumngrid
\appendix

\section{Dependence on hidden layer size}\label{app:A}

In this appendix, we examine how the performance of RBM training scales with the hidden-layer size $N_H$, comparing classical persistent contrastive divergence (PCD) and diabatic quantum annealing (DQA). We focus on convergence behavior, validation error, and the resulting performance gap as the model dimension increases, providing a systematic characterization of the system-size dependence of DQA within the range accessible to current hardware.

\subsection{Exponential fitting of validation error versus hidden-layer size}
\begin{figure}[h!]
    \centering
    \includegraphics[width=1.0\linewidth]{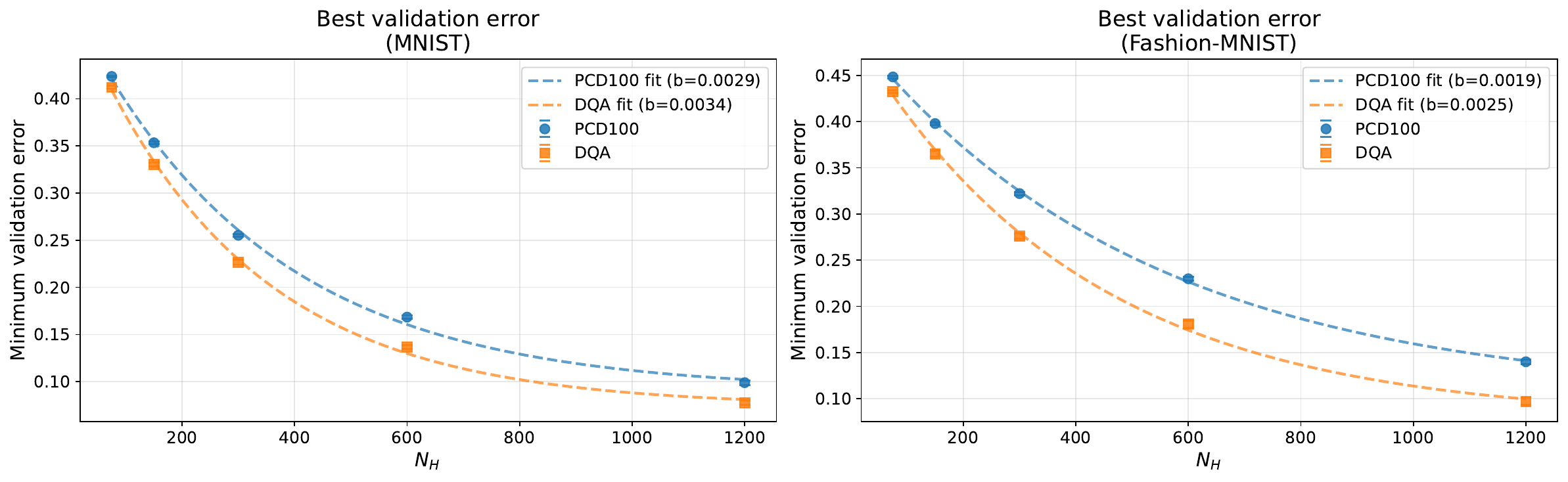}
    \caption{Minimum validation error as a function of the hidden-layer size $N_H$ for MNIST and Fashion-MNIST. Mean minimum validation error across 10 independent runs is shown for PCD and DQA. Dashed lines indicate fits to the exponential form $y = a e^{-b N_H} + c$. While both methods exhibit exponential decay with increasing model capacity, DQA consistently attains a larger decay rate $b$, indicating better scalability.}
    \label{fig:min_val}
\end{figure}

\begin{figure}[h!]
    \centering
    \includegraphics[width=1.0\linewidth]{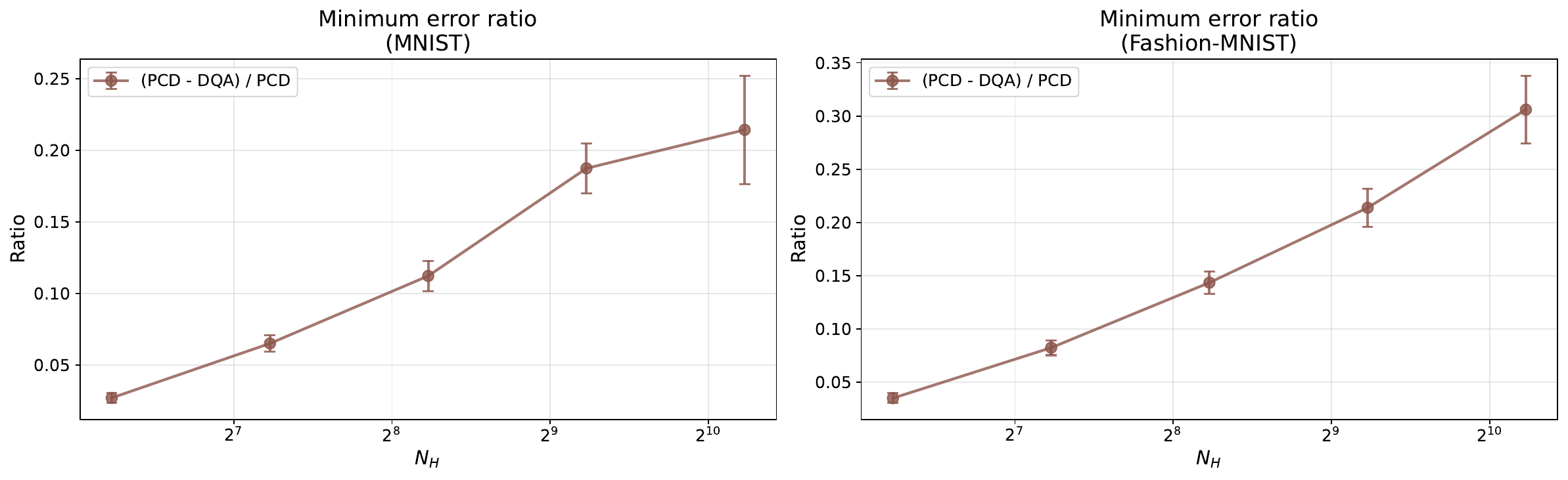}
    \caption{Relative improvement in the minimum validation error as a function of the hidden layer size $N_H$, defined as $(\mathrm{PCD} - \mathrm{DQA})/\mathrm{PCD}$, for MNIST and Fashion-MNIST. The increasing ratio with system size indicates that the performance advantage of DQA over PCD becomes more pronounced as the model dimension grows.}
    \label{fig:gap_ratio}
\end{figure}

Figure~\ref{fig:min_val} shows the minimum validation error as a function of the hidden-layer size $N_H$ for both MNIST and Fashion-MNIST. For each method, the data are well described by an exponential form
\begin{equation}
y = a e^{-b N_H} + c,
\end{equation}
the fitting parameters are as follow: $a=0.4125, b=0.0029, c=0.0902$ (PCD, MNIST), $a=0.4330, b=0.0034, c=0.07373$ (DQA MNIST), $a=0.3978, b=0.0019, c=0.1022$ (PCD Fashion-MNIST), $a=0.4219, b=0.0025, c=0.0774$ (DQA Fashion-MNIST).
The decay rate $b$ is approximately 20 percent larger for DQA. As the system size increases, the minimum validation error decreases exponentially, with the decay rate for DQC being noticeably steeper. 

Figure~\ref{fig:gap_ratio} further quantifies this trend by plotting the relative improvement in the minimum validation error, defined as $(\mathrm{PCD} - \mathrm{DQA})/\mathrm{PCD}$. The ratio increases monotonically with $N_H$, demonstrating that the performance gap between the two methods widens as the system size grows. This behavior suggests that the advantage of DQA becomes more pronounced in larger models.

\subsection{Exponential fitting of learning curves}
\begin{figure}[h!]
    \centering
    \includegraphics[width=1.0\linewidth]{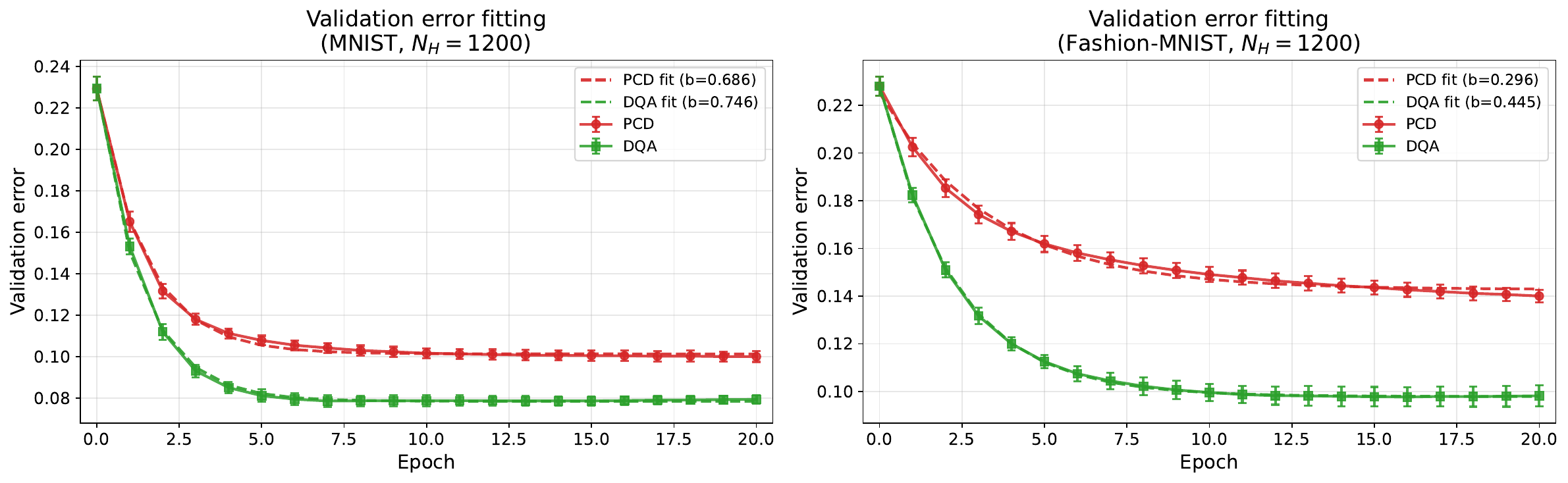}
    \caption{Validation error as a function of training epoch for $N_H = 1200$ on MNIST and Fashion-MNIST. Data points represent the mean over 10 independent runs. Solid lines show exponential fits of the form $y = a e^{-b t} + c$. DQA exhibits a larger decay parameter $b$, indicating faster convergence compared to PCD.}
    \label{fig:fitting}
\end{figure}

\begin{figure}[h!]
    \centering
    \includegraphics[width=1.0\linewidth]{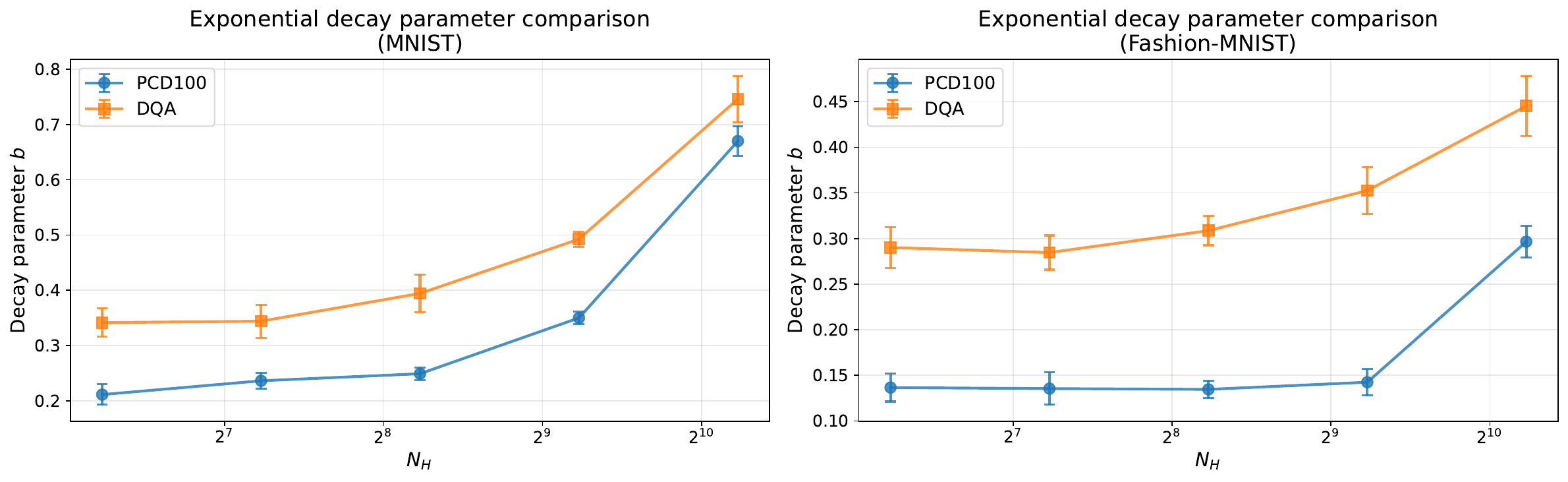}
    \caption{Comparison of the exponential decay parameter $b$ as a function of the hidden-layer size $N_H$ for MNIST and Fashion-MNIST. Error bars indicate the standard deviation across 10 independent runs. For all tested system sizes, DQA yields consistently larger decay rates than PCD, demonstrating that the convergence-speed advantage persists with increasing model dimension.
}
    \label{fig:hidden_size}
\end{figure}

To isolate the effect of training speed, Fig.~\ref{fig:fitting} shows the validation error as a function of training epoch for a representative large system ($N_H = 1200$). The learning curves are well captured by an exponential decay in epoch,
\begin{equation}
y = a e^{-b \cdot\mathrm{Epoch}} + c,
\end{equation}
where the decay parameter $b$ characterizes the convergence rate and $c$ denotes the baseline of the validation error. The fitting parameters are as follow: $a=0.128, b=0.686, c=0.101$ (PCD MNIST), $a=0.152, b=0.746, c=0.078$ (DQA MNIST), $a=0.082, b=0.296, c=0.143$ (PCD Fashion-MNIST),
$a=0.130, b=0.445, c=0.098$ (DQA Fashion-MNIST). In both datasets, the fitted value of $b$ is larger for DQA than for PCD, indicating faster convergence under DQA-based sampling.

Figure~\ref{fig:hidden_size} extends this analysis by comparing the decay parameter $b$ across different hidden-layer sizes. For all tested $N_H$, the convergence rate remains systematically higher for DQA. Importantly, the separation between the two methods does not diminish with increasing system size, indicating that the speedup provided by DQA is not a finite-size effect.

\begin{figure}[h!]
    \centering
    \includegraphics[width=1.0\linewidth]{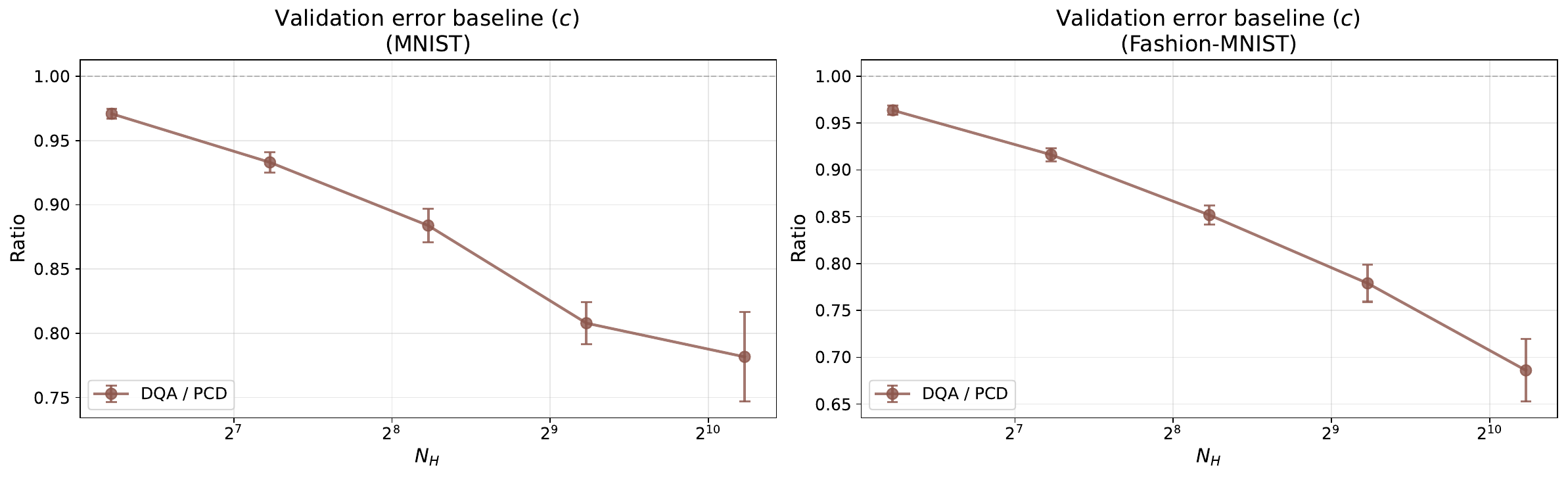}
    \caption{Ratio of the asymptotic validation error baselines $c_{\mathrm{DQA}}/c_{\mathrm{PCD}}$ as a function of the hidden-layer size $N_H$ for MNIST and Fashion-MNIST. Since $c$ represents the long-time validation error floor, the decreasing ratio with increasing $N_H$ indicates that DQA achieves systematically better final solutions as the system size grows.}
    \label{fig:baseline}
\end{figure}

Figure~\ref{fig:baseline} examines the ratio of the asymptotic baseline parameters $c$ extracted from the exponential fits. The ratio $c_{\mathrm{DQA}}/c_{\mathrm{PCD}}$ decreases with increasing hidden-layer size for both datasets. Since $c$ represents the validation-error floor reached in the long-training limit, this trend indicates that DQA not only accelerates learning but also achieves a lower limiting error. Because both methods train identical model architectures, the reduced baseline is naturally attributed to improved sampling quality under DQA, which yields more accurate gradient estimates and, consequently, better-trained models.

\subsection{Summary}

Taken together, these results demonstrate that the advantage of DQA-based training manifests in three complementary ways: (i) faster convergence, (ii) lower asymptotic validation error, and (iii) an increasing performance gap with system size. These trends are consistent with the analytical sampling-time considerations discussed in Section~\ref{app:B}, where classical MCMC methods suffer from growing correlation times as the system size increases, whereas DQA generates decorrelated samples in a single annealing run.

\section{System-size scaling of sampling cost}\label{app:B}
We analyze the system-size scaling of the sampling cost in PCD and DQA, focusing exclusively on the sample-generation stage prior to parameter updates. We consider a dense RBM with $N_v$ visible units and $N_h$ hidden units, and define the total system size as $N = N_v + N_h$. The number of trainable couplings scales as $\mathcal{O}(N^2)$.
In PCD, samples are generated via alternating conditional updates between the visible and hidden layers. A single Gibbs step consists of one visible-to-hidden and one hidden-to-visible transition, each involving weighted sums over all couplings. The computational cost of one Gibbs step therefore scales as
\begin{equation}
t_{\mathrm{Gibbs}} \in  \mathcal{O}(N^2).
\end{equation}

To obtain approximately independent samples, multiple Gibbs steps are required to reduce correlations between successive states. Denoting the required number of steps by $K(N)$, the sampling cost per independent sample scales as
\begin{equation}
t_{\mathrm{PCD}} \sim K(N)\, t_{\mathrm{Gibbs}}.
\end{equation}
Due to dynamical decorrelation effects, $K(N)$ increases with system size and is commonly associated with critical slowing down, $K(N)\sim N^{z}$ with $z>0$. This leads to
\begin{equation}
t_{\mathrm{PCD}} \in \mathcal{O}(N^{2+z}).
\label{eq:t_pcd_scaling}
\end{equation}

In DQA-based sampling, a single sample is obtained from one annealing run. The total cost per sample can be decomposed as
\begin{equation}
t_{\mathrm{DQA}} = t_{\mathrm{prep}}(N) + \tau + t_{\mathrm{readout}},
\label{eq:t_dqa_decomposition}
\end{equation}
where $t_{\mathrm{prep}}(N)$ denotes the state-preparation time required to program the model parameters in hardware level, $\tau$ is the annealing time and $t_{\mathrm{readout}}$ is the measurement time. For dense models, state preparation scales with the number of couplings,
\begin{equation}
t_{\mathrm{prep}} \in  \mathcal{O}(N^2),
\label{eq:t_prep}
\end{equation}
while both $\tau$ and $t_{\mathrm{readout}}$ can be kept approximately constant. Consequently, the total DQA sampling cost scales as
\begin{equation}
t_{\mathrm{DQA}} \in \mathcal{O}(N^2)+\mathcal{O}(1)\sim \mathcal{O}(N^2).
\label{eq:t_dqa_scaling}
\end{equation}

Equations~(\ref{eq:t_pcd_scaling}) and (\ref{eq:t_dqa_scaling}) highlight the qualitative difference between the two approaches: classical sampling incurs an additional multiplicative overhead associated with dynamical decorrelation, whereas DQA replaces this time-domain overhead with a purely static state-preparation cost.
\[
\boxed{
\begin{aligned}
t_{\mathrm{PCD}} &\in \mathcal{O}(N^{2+z}) \text{ with } z>0, \\
t_{\mathrm{DQA}} &\in \mathcal{O}(N^{2})
\end{aligned}
}
\]

\section{Annealing schedule}\label{app:C}
\begin{figure}[h!]
    \centering
    \includegraphics[width=0.65\linewidth]{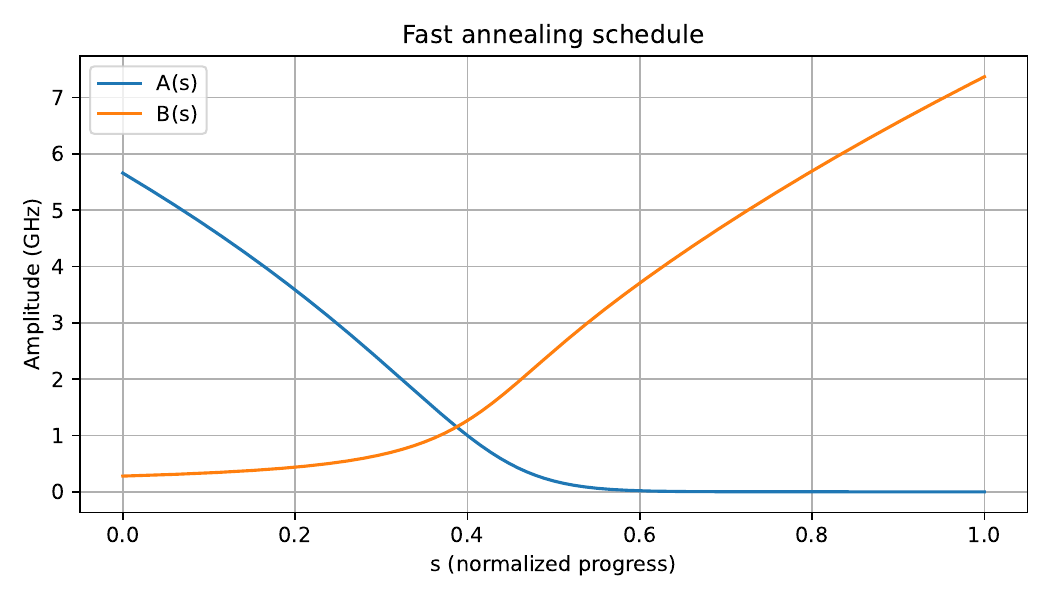}
    \caption{Default annealing schedule of the D-Wave Advantage2 system1-5 in fast annealing mode~\cite{dwave-schedule}.  
    This schedule was fixed by the hardware and used for all DQA experiments.  
    Only the annealing time was varied, and an annealing time of $5\,\mathrm{ns}$ was found to yield an effective inverse temperature close to $\beta=1$.}
    \label{fig:dwave_schedule}
\end{figure}

For DQA, two items must be specified: (i) the annealing schedule ($A(s), B(s)$) and (ii) the annealing time ($\tau$).  
On the D-Wave platform, when operating in fast annealing mode, the schedule cannot be modified and the default schedule shown in Fig.~\ref{fig:dwave_schedule} was used. We therefore explored only the annealing time. The results indicate that an annealing time of $5\,\mathrm{ns}$ yields an effective inverse temperature close to $\beta\simeq1$.  

Finer tuning would have been possible if fractional annealing times were supported, but this option was not available. In principle, even if the annealing time is set differently, one can compensate by rescaling the energy according to the effective temperature corresponding to that schedule and time. However, given that the resolution of the programmable couplings $J$ on the current hardware is about $0.01$~\cite{dorband2018extending}, it is preferable to choose annealing times that naturally align with $\beta \simeq 1$. Also, one must also take into account the coherence time of the hardware. In our case, 5 ns lies within the regime that the D-Wave machine is claimed to maintain coherence~\cite{dwave_coherence,dwave_report}.

\end{document}